\documentclass[conference]{IEEEtran}
\IEEEoverridecommandlockouts
\usepackage{authblk}
\usepackage{multirow}
\usepackage[normalem]{ulem}
\useunder{\uline}{\ul}{}
\usepackage{placeins}
\usepackage{float}
\usepackage{caption}
\usepackage{subcaption}
\usepackage{array}
\usepackage{multirow}
\usepackage{cite}
\usepackage{amsmath,amssymb,amsfonts}
\usepackage{algorithmic}
\usepackage{graphicx}
\usepackage{textcomp}
\usepackage{xcolor}
\usepackage{tabularx}

\newcolumntype{P}[1]{>{\centering\arraybackslash}p{#1}}
\newcolumntype{M}[1]{>{\centering\arraybackslash}m{#1}}
\makeatletter
\newcommand*{\rom}[1]{\expandafter\@slowromancap\romannumeral #1@}
\makeatother
\usepackage{cleveref}
\graphicspath{ {./Figures/} }
\def\BibTeX{{\rm B\kern-.05em{\sc i\kern-.025em b}\kern-.08em
    T\kern-.1667em\lower.7ex\hbox{E}\kern-.125emX}}
    
\begin{document}

\title{A New Multimodal Medical Image Fusion based on Laplacian  Autoencoder with Channel Attention

% Multimodal Medical Image Fusion using  Convolutional Autoencoder with Channel Attention Weighted Pooling
}

\author{Payal Wankhede, Manisha Das,~\IEEEmembership{Student Member,~IEEE}, Deep Gupta, ~\IEEEmembership{Senior Member,~IEEE,}  \\Petia  Radeva, ~\IEEEmembership{Fellow,~IAPR,} and Ashwini M Bakde
        % <-this % stops a space
        % <-this % stops a space        % <-this % stops a space
\thanks{Payal Wankhede is with the Department
of Electronics and Comm. Engineering, Visvesvaraya National Institute of Technology Nagpur 40010, India. e-mail: (payallaptopid@gmail.com).}
\thanks{Manisha Das is with the Department
of Electronics and Comm. Engineering, Visvesvaraya National Institute of Technology Nagpur 40010, India. e-mail: (das.manisha1989@gmail.com ).}% <-this % stops a space
\thanks{Deep Gupta is with the Department
of Electronics and Comm. Engineering, Visvesvaraya National Institute of Technology Nagpur, 440010, India. e-mail: (deepgupta@ece.vnit.ac.in).}
\thanks{Petia Radeva is with the Department de Mathematics and Informatics, Universitat de Barcelona, 08007 Barcelona, Spain, and also with the Computer Vision Center, 08193 Cerdanyola, Spain. e-mail: (petia.ivanova@ub.edu).}
\thanks{Ashwini Bakde is with the Department of Radio-Diagnosis, All India Institute of Medical Sciences Nagpur, 441108, India. e-mail: (ashwini@aiimsnagpur.edu.in).}

\thanks{Manuscript received XX, 2023. }}

\maketitle

\begin{abstract} 
Medical image fusion combines the complementary information of multimodal medical images to assist medical professionals in the clinical diagnosis of patients' disorders and provide guidance during preoperative and intra-operative procedures. Deep learning (DL) models have proved to achieve end-to-end image fusion with high robust and accurate fusion performance. However, most of the DL-based fusion models  perform down-sampling on the input images to minimize the number of learnable parameters and computations. During this process, salient features of the source images become irretrievable leading to the loss of crucial diagnostic edge details and contrast of various brain tissues. In this paper, we propose a new  multimodal medical image fusion model is proposed that is based on integrated  Laplacian-Gaussian concatenation with  attention  pooling (LGCA). We prove that our model preserves  effectively complementary information and important tissue structures. 
% Extensive experimentation is carried out to apprehend the qualitative and quantitative performance of the proposed fusion method. 
Extensive experimental results demonstrate a notable improvement in the fusion performance of the proposed method compared to the recently developed state-of-the-art fusion approaches on 4 different image modalities using 6 different statistics metrics.
\end{abstract} 


\section{Introduction}
Multimodal medical image fusion plays an important role in extracting and integrating the complementary details from source modalities to achieve more comprehensive representation and factual conclusions \cite{li_pixel-level_2017, singh2019, das2022tim}. The source imaging modalities can include anatomical information from magnetic resonance imaging (MR), computed tomography (CT), as well as functional information from positron emission tomography (PET) or single-photon emission computed tomography (SPECT) \cite{du_overview_2016, 9261602}. The MR images capture the soft tissue structures with high spatial resolution, while the CT images accurately detect the bone structures, however not able to reflect the soft tissue contrast. whereas, SPECT and PET scans identify changes in metabolic functions and regional chemical composition, generating rich color information but have a relatively low resolution compared to MR images. Multimodal image fusion allows the visualization of contrasting information in a single fused image, thereby assisting in image-guided interventions and invasive procedures, achieving a superior medical diagnosis \cite{hermessi_multimodal_2021}.

Many different methods based on multi-scale decomposition \cite{du2019intrinsic, singh2019multimodal, liu2018multi, li2020laplacian, yin2018medical}, sparse representation \cite{zong2017medical, wang2020multimode}, fuzzy logic \cite{yang2016multimodal, manchanda2018improved}, support vector machines \cite{padmavathi_fusion_2017, zhang_multi-kernel_2009}, etc. have been used to accomplish the medical image fusion. However, these methods present significant hurdles because of using conventional fusion rules, hand-crafted features, manual intervention in selection the decomposition levels, etc., limiting the overall fusion performance \cite{kaur_image_2021}. In recent years, deep learning (DL) has made significant strides in many computer vision and image processing challenges, including classification, segmentation, and fusion \cite{cai2020review, liu_deep_2018}. In the case of supervised DL algorithms, models such as convolutional sparse representation \cite{liu_medical_2019, liu2020medical}, auto-encoders \cite{wang2018image, tawfik2022multimodal} and convolutional neural networks (CNN) \cite{wang2020multi, xia2019novel, singh2019multimodal} have elevated the fusion performance as compared to the conventional fusion methods \cite{zhang2021image}. However, these models have a limited ability to preserve the contrast and structural details of the source images and are confined to the image dataset on which they are trained. Additionally, the use of supervised DL models is challenging due to the absence of ground truth for fusion in case of medical images.  The unsupervised DL models such as generative adversarial networks (GANs) \cite{ma_ddcgan:_2020, fu_dsagan:_2021} and dense convolutional networks \cite{zhang_sdnet:_2021, xu_u2fusion:_2020} based fusion techniques lessen the need for ground truth. However, some crucial diagnostic information may not be retained while extracting the characteristic information present in the source images using the conventional model architectures and  loss functions.

Recently, Convolutional autoencoder (CAE) based image fusion methods have also been reported with superior fusion performance \cite{azarang2019convolutionalae, qu2021mssl}. CAE is an unsupervised dimensionality reduction model made up of convolutional layers that can generate compressed image representations \cite{zhang_better_2018}. They are used to minimize reconstruction errors while performing image reconstruction. This is achieved by learning the optimal convolutional filters. Once trained, they can be utilized to extract the features from any input dataset \cite{masci_stacked_2011},\cite{chen_deep_2017}. However, in the conventional CAE models the direct downsizing of the input features using pooling layers results in aliasing-induced distortion \cite{vasconcelos_impact_2021}. Aliasing causes the high-frequency components to become indistinguishable from the low-frequency components \cite{ribeiro2021convolutional}. For edge preservation, high-frequency information is vital. Although anti-aliasing convolutional neural networks (ACNN) have been modeled to reduce the aliasing effect, they employ max pooling operation, which generates an unstable output in case of affine transformations \cite{azulay_why_2018}.  Average and max pooling are two popular pooling techniques used at the down-sampling stage in most of the DL models. The advantage of averaging is that it minimizes the impact of noisy features. However, since it assigns equivalent weights to all elements in the pooling kernel, background characteristics gain dominance in the resulting down-sampled representation, which leads to reduced discriminating power. In case of max pooling, the highest pixel value in each pooling region is selected, avoiding the effect of undesired background information. Due to this, the down-sampled representation may capture noisy features, and the loss of salient features is also observed in the further stages of model\cite{nirthika2022pooling}. 

% To address the above-discussed challenges, this paper  proposes a medical image fusion method that employs the laplacian-gaussian concatenation with attention pooling discussed in \cite{sineesh_exploring_2021}. With the aforementioned constraints in mind, the proposed fusion technique utilizes a convolutional autoencoder framework, which integrates the LGCA pooling model at the down-sampling stage. In LGCA pooling, an attention mechanism is performed on the extracted laplacian and gaussian filter components. The areas of sharp changes with high frequency are captured by the laplacian filter components. The low frequencies, which contain most of the color data along with the overall spatial information of the image, are detected by gaussian filter components. The CAE learns the crucial features of the source images by compressing the image data in a single vector representation and performing the subsequent reconstruction. The model is further trained and tested with the objective of executing medical image fusion. 

To address the above-discussed challenges, this paper proposes a novel multimodal medical image fusion method based on convolutional autoencoder that employs the Laplacian-Gaussian concatenation with attention (LGCA) pooling. It integrates the LGCA pooling layer at the down-sampling to retain both the structural and textural details of the extracted feature maps. In LGCA pooling, an attention mechanism is performed on the extracted Laplacian and Gaussian filter components \cite{sineesh_exploring_2021}. The areas of sharp changes with high frequency are captured by the Laplacian filter components and the low frequencies, which contain  overall spatial information of the image, are detected by Gaussian filter components. The attention mechanism which is performed on the concatenated channels is based on a squeeze and excitation (SE) network \cite{hu_squeeze-and-excitation_2018}. By estimating the weights for each channel, the SE network assists in emphasizing the significant and relevant features in the input. The CAE model learns the crucial features of the source images by compressing the image data into a single vector representation and performing the subsequent reconstruction. 

The rest of the paper is organized as follows, Section \ref{section:2} discusses the details of LGCA pooling and squeeze and excitation network used in the proposed fusion approach. Section \ref{section:3} presents in detail the overview and steps of the proposed fusion model and its architecture. Section \ref{section:4} discusses the implementation and parameter settings. In Section \ref{section:5}, visual and quantitative performance analysis and the ablation study with conventional pooling methods are presented to verify the effectiveness of the proposed method over existing fusion methods followed by the conclusions in Section \ref{section:6}.
 
\section{Methodology} \label{section:2}
\subsection{Laplacian-Gaussian Concatenation with Attention Pooling} 

 The objective of LGCA pooling is to preserve low and high-frequency structural and textural details of the input image with the help of a channel attention mechanism \cite{sineesh_exploring_2021}. 
% Therefore, before implementing the channel attention mechanism, extract and concatenate the low and high-frequency components of the input image. 
A Gaussian filter is applied to the extracted input feature maps and  the resulting low-frequency components are subtracted from the input feature maps to acquire high-frequency (Laplacian) components. Both filter components are then concatenated, as a result, the number of channels is doubled after concatenation. Consider an image $X$, composed of dimensions $H \times W \times C_{in}$, where the height and width of the image are represented by $H$ and $W$, respectively, and $ C_{in}$ represents the number of channels. The concatenation is performed as follows,
\begin{equation} \label{equation2}
C\{F(X)\} = \{ G(X), F(X) - G(X) \} 
\end{equation} 
where the $F(X)$ denotes the extracted features of the input image, \begin{math} G(X) \end{math} and  \begin{math} F(X) - G(X) \end{math} represents the Gaussian filter component and the Laplacian filter component, respectively.

\begin{figure}[!t] 
\centering
	\scalebox{0.45}{\includegraphics[width=7.4in, %height=3.4in
		]{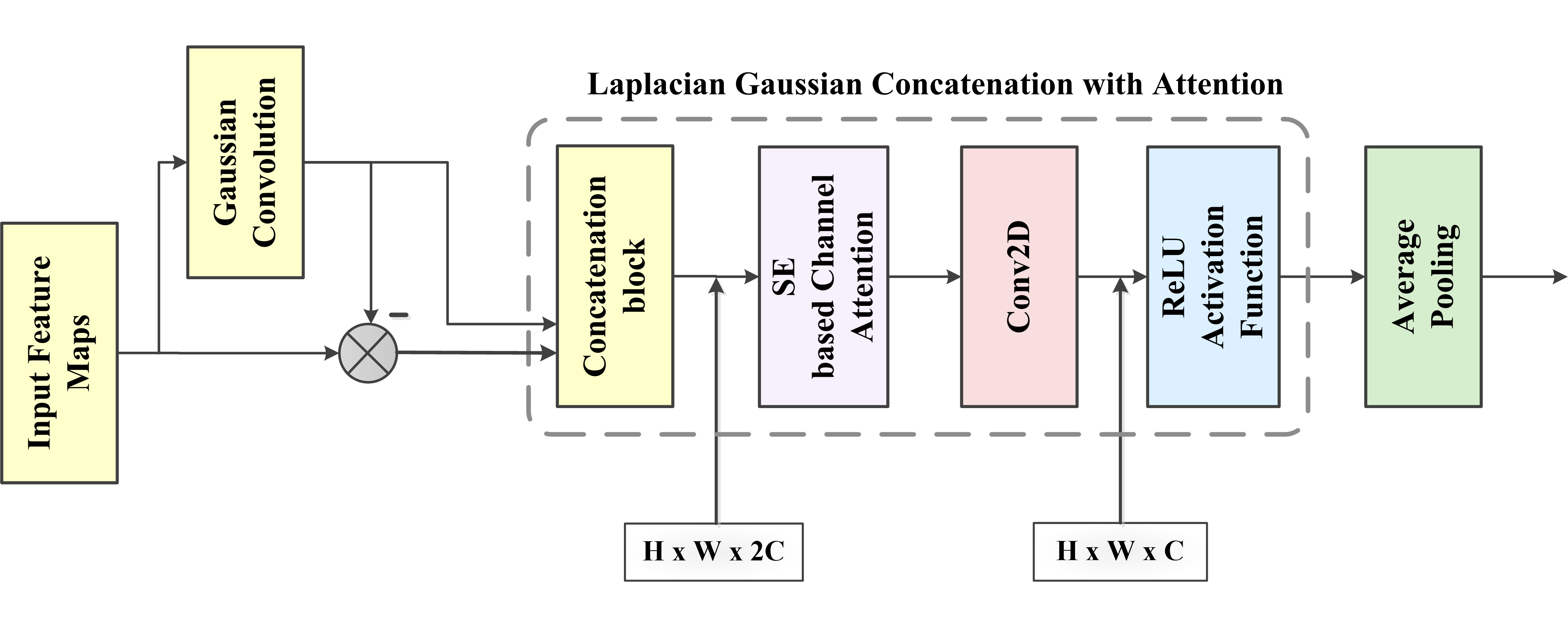}} 
\caption{Laplacian Gaussian concatenation with attention pooling.} 
\label{fig:LGCA} 
\end{figure}

The block diagram for the LGCA pooling technique is shown in  \figurename{~\ref{fig:LGCA}}. The concatenated feature maps $C\{F(X)\} $ are passed through the channel attention block where attention-weighted feature maps are generated using a squeeze and excitation (SE) network. The output feature dimensions after channel attention remain $H \times  W \times  2C$. A convolution operation is performed for restoring the original dimensions of the feature map, which ensures that the aggregation of channel dimensions is done according to the application. It is followed by a non-linear ReLU activation function. An average pooling operation is utilized on the attention-weighted feature maps of size $H \times  W \times  C$ to summarize the feature attributes and also provides translation invariance to the attention-weighted output.
%%%%%%%% Figure  LGCA Pooling Image %%%%%%%  

% %%%%%%%% Figure  LGCA Pooling Image %%%%%%%  
% \begin{figure}[hbt!] 
% \centerline{\includegraphics[width=.5\textwidth, height=.22\textwidth]{LGCA block diagram image}} 
% \caption{Laplacian Gaussian Concatenation with Attention.} 
% \label{fig:LGCA} 
% \end{figure}
% \FloatBarrier
%\vspace{0.5cm}

\subsection{The Squeeze and Excitation Network} 

\begin{figure}[!t] 
\centering
	\scalebox{0.4}{\includegraphics[width=7.4in, %height=3.4in
		]{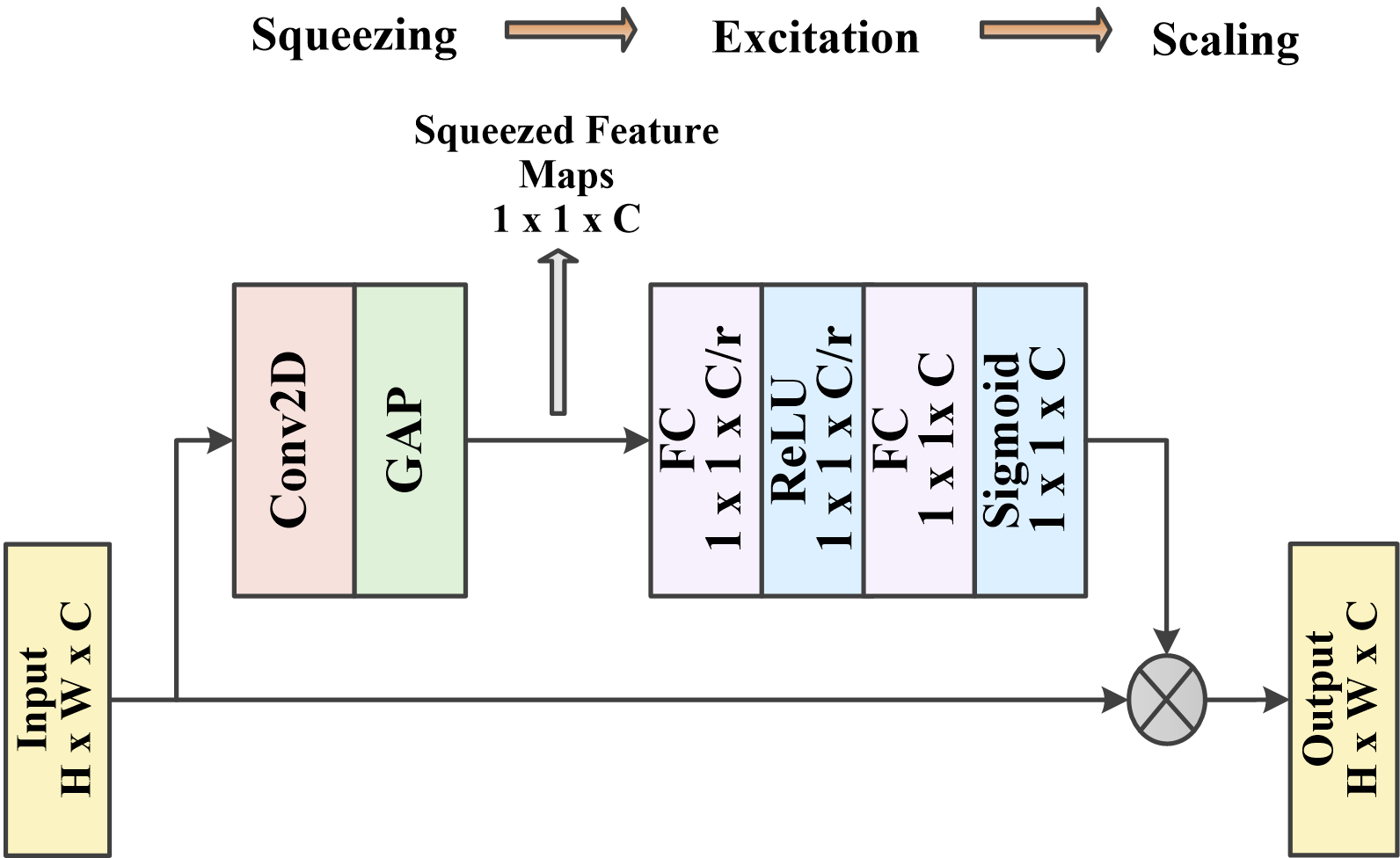}} 
\caption{The squeeze and excitation function.}  \label{fig:SE-func} 
\end{figure}

A crucial part of the LGCA pooling technique is the channel attention block which performs the squeeze and excitation (SE) function. The SE network \cite{hu_squeeze-and-excitation_2018}  re-weights each channel appropriately, making it more sensitive to significant features while discarding irrelevant elements. It performs three operations on the input image squeezing, excitation, and scaling. \figurename{~\ref{fig:SE-func}} shows the architecture of the SE network.

Consider an image with dimension $H \times W \times C_{in}$, where the height and width of the image are represented by $H$, $W$, respectively, and $ C_{in} $ represents the original number of channels. After a simple convolution operation, a feature map of size \begin{math} H \times W \times C \end{math} with different numbers of channels \begin{math} C \end{math} is extracted from the input. To extract the global information from each channel of the feature map, the squeeze operation is performed. This is achieved by implementing global average pooling (GAP) on the feature map to reduce the spatial dimension from $H \times W \times C$ to $1 \times 1 \times C $.

%%%%%%%% Figure  Squeeze and Excitation Image %%%%%%% 

The excitation operation is performed on these single descriptors for each channel. Two fully connected layers with a bottleneck architecture surrounding the non-linear ReLU activation function are utilized for the excitation operation. This serves to determine the per-channel weights, which are used to adaptively re-scale the input feature map. In this bottleneck architecture, the input dimensionality-reduction layer reduces the number of channels by a factor \begin{math} r=16 \end{math}, which is later restored to the original number by the output dimensionality-increasing layer. The sigmoid activation function is used for estimating per-channel attention weights for each squeezed embedding. A scaling operation is performed by applying an element-wise multiplication between the output of sigmoid activation function and the input feature map. The sigmoid activation function assigns the values from $0$ to $1$ to the squeezed embedding, which helps in providing weights and generating channels with appropriate importance.

\section{Proposed Method} \label{section:3}
This section discusses in detail the network architecture of the LGCA attention-pooled CAE and the framework of the proposed fusion method.

\subsection{The Network Architecture} \label{model_arch}
\begin{figure}[!t]
\centering
	\scalebox{0.45}{\includegraphics[width=7.4in, %height=3.4in
		]{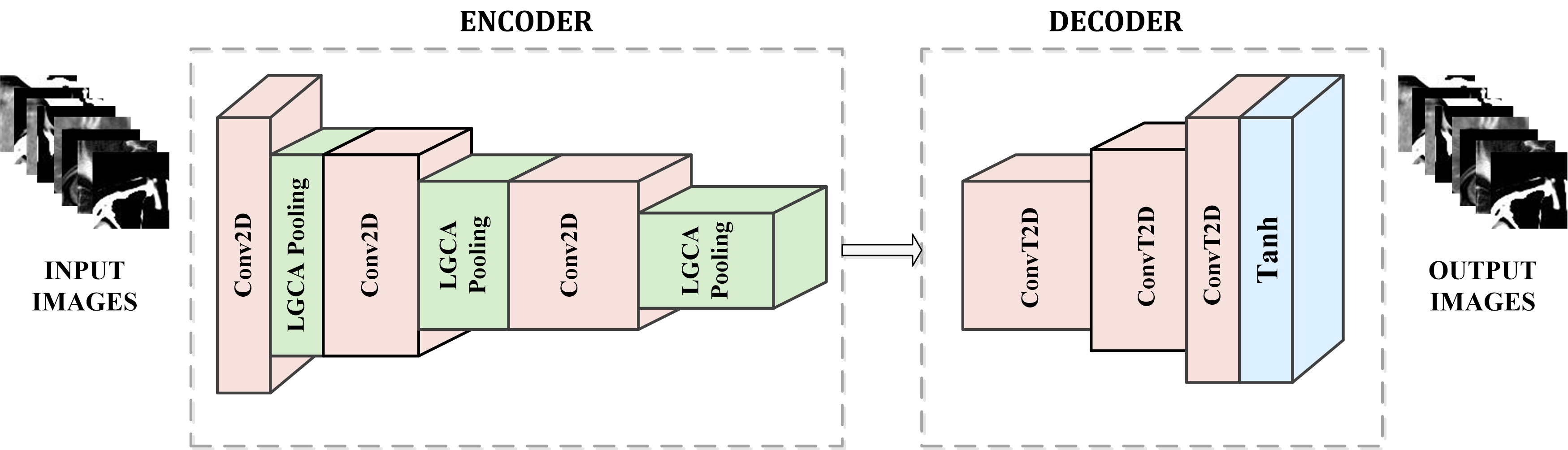}}
\caption{Network architecture of the LGCA pooled CAE used in the proposed fusion method}
\label{fig:ModelArchitecture}
\end{figure}

The network architecture of the LGCA attention-pooled CAE is 
shown in \figurename{~\ref{fig:ModelArchitecture}}.
%\FloatBarrier
The proposed training model embodies a convolutional autoencoder DL architecture. The input is an image of dimension \begin{math} 256 \times 256 \end{math}, which is passed through an encoder and the reconstruction of this image is obtained at the output of decoder. The encoder block consists of three convolutional layers and three LGCA pooling layers comprised in an alternate manner. The convolution layer is a 2-D convolution, where a kernel of fixed size $3 \times 3$ slides over the 2-D input data with stride = 1, executing element-wise multiplication and summing the result into a single output pixel. It performs the task of extensive feature extraction. The LGCA pooling serves in preserving both the low and high-frequency characteristics of the extracted feature maps. The pooling layers also execute dimensionality reduction from spatial size \begin{math} 256 \times 256 \end{math} to the final output of the encoder with dimension \begin{math} 32 \times 32 \end{math}. Three transposed convolution layers, which perform deconvolution of the encoded input, are included in the decoder block to obtain the reconstructed image. A transposed convolutional layer aims to reconstruct the spatial dimensions of the convolutional layer and reverses the down-sampling techniques applied to it. The kernel size for these layers is fixed to $2 \times 2$ with stride = 2. A $tanh$ activation function is used at the last stage of the decoder. This completes the training model architecture of the proposed method. This model is trained on a medical dataset containing numerous neurological images. The trained model is utilized for the execution of multimodal medical image fusion. \tablename{~\ref{table:ModelArch_table}} summarizes the number of input and output channels of the convolutional layers used in encoder and decoder blocks shown in \figurename{~\ref{fig:ModelArchitecture}}.

%%%%%%%% Figure Model Architecture Image %%%%%%% 

%%%%%%%% Model Architecture Table %%%%%%%
\begin{table}[!t]
\centering
\caption{Channel details for the proposed training model}
\label{table:ModelArch_table}
\begin{tabular}{llllllllllll}
\cline{1-4}
\multicolumn{1}{c}{Model} & \multicolumn{1}{c}{Layer} & Input Channel & Output Channel \\ \cline{1-4}
\multirow{3}{*}{Encoder} & \multicolumn{1}{c}{Conv2D (1)} & \multicolumn{1}{c}{1} & \multicolumn{1}{c}{64} \\
 & \multicolumn{1}{c}{Conv2D (2)} & \multicolumn{1}{c}{64} & \multicolumn{1}{c}{128} \\
 & \multicolumn{1}{c}{Conv2D (3)} & \multicolumn{1}{c}{128} & \multicolumn{1}{c}{256} \\ \cline{1-4}
\multicolumn{1}{c}{\multirow{3}{*}{Decoder}} & \multicolumn{1}{c}{ConvT2D (1)} & \multicolumn{1}{c}{256} & \multicolumn{1}{c}{128} \\
\multicolumn{1}{c}{} & \multicolumn{1}{c}{ConvT2D (2)} & \multicolumn{1}{c}{128} & \multicolumn{1}{c}{64} \\
\multicolumn{1}{c}{} & \multicolumn{1}{c}{ConvT2D (3)} & \multicolumn{1}{c}{64} & \multicolumn{1}{c}{1} \\ \cline{1-4}
\end{tabular}
\end{table}
\FloatBarrier
\subsection{Proposed Fusion Framework}

%%%%%%%% Figure Fusion Strategy Image %%%%%%%  hbt!
\begin{figure*}[hbt!]
\centering
	\scalebox{0.7}{\includegraphics[width=7.4in, %height=3.4in
		]{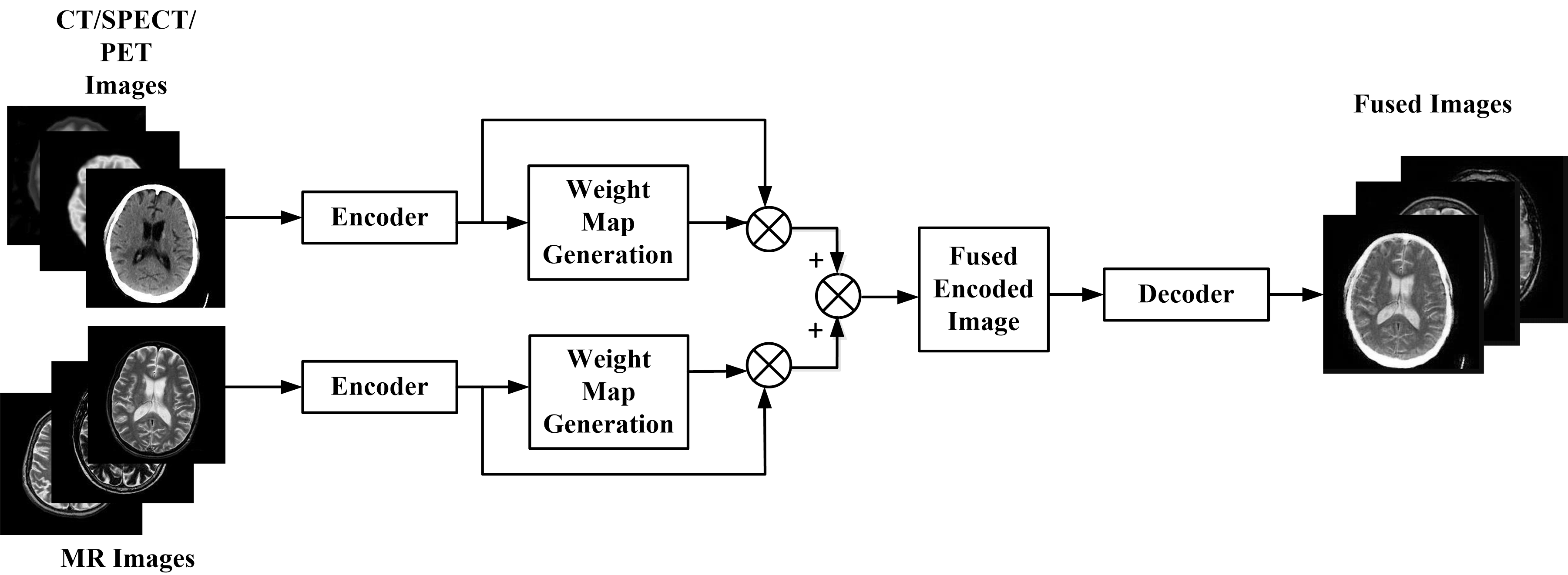}}
\caption{Image fusion framework using the trained model.}
\label{fig:FusionFramework}
\end{figure*}
%\FloatBarrier

The framework for the proposed fusion method is shown in \figurename{~\ref{fig:FusionFramework}}. The trained CAE model discussed in the previous section \ref{model_arch} is utilized for the fusion of multimodal medical images. The two source images are passed through the encoder of the trained model and their encoded feature maps are obtained. Weight maps are obtained using a weight map generation block that captures the feature sparsity of each of the encoded feature maps. The generated weights are applied to the individual encoded source images. Next, a summation of the two weighted-encoded inputs is performed, and the encoded fused image is obtained. The encoded version of the fused image is then passed through the decoder which reconstructs the final fused image. 

The implementation steps of the fusion strategy are as follows:\\

\textbf{Step 1:} Consider a pair of pre-registered CT/SPECT/PET and MR images represented as $P = P_{i,j}$ and $Q = Q_{i,j}$, respectively, where $i$ and $j$ represent the row and column indices. For color images (SPECT/PET), first, convert the images from \textit{RGB} to \textit{YUV} color space using the following equation; 
\begin{equation} 
\begin{bmatrix} 
Y_c \\
U_c \\
V_c \\
\end{bmatrix} = 
\begin{bmatrix} 
0.299 & 0.587 & 0.114 \\
$- 0.169 $ & $- 0.331$ & 0.5 \\
0.5 & $- 0.419$ & $- 0.081$ \\
\end{bmatrix}
\begin{bmatrix} 
R_c \\
G_c \\
B_c \\
\end{bmatrix}
\end{equation}
Where, $R_c$, $G_c$, and $B_c$ refer to the red, green, and blue color channels of an image and  $Y_c$, $U_c$, and $V_c$, represent the one luminance and two chrominance components, respectively. Now, consider the luminance component $Y_c$  for further processing.\\

\textbf{Step 2:} Subject the source images, $P_{i,j}$ and $Q_{i,j}$ to the encoder ($E$) of the trained CAE model and get encoded features $E^P$ and $E^Q$ as follows;
\begin{equation}
    E^P = E(P_{i,j})
\end{equation}
\begin{equation}
    E^Q = E(Q_{i,j})
\end{equation} 

\textbf{Step 3:} In order to retain only the sparse features, an  activity map $AM_{ij}$ is generated using $l_1$-norm of the encoded features $E^P$ and $E^Q$ as follows;
\begin{equation}
    AM^{P}_{ij} 
    =
    \begin{Vmatrix}
       E^P 
    \end{Vmatrix}_1
\end{equation}
\begin{equation}
    AM^{Q}_{ij} 
    =
    \begin{Vmatrix}
       E^Q 
    \end{Vmatrix}_1
\end{equation}

\vspace{0.3cm}

\textbf{Step 4:} A final activity map $FAM_{ij}$ is generated for each $AM_{ij}$ by performing block based averaging within a window of $3 \times 3$ to improve the robustness towards misregistration. Next, the weight maps are generated as follows;
\begin{equation}
    W^{P}_{ij} = \frac{FAM^{P}_{ij}}{FAM^{P}_{ij} + FAM^{Q}_{ij}}
\end{equation}
\begin{equation}
    W^{Q}_{ij} = \frac{FAM^{Q}_{ij}}{FAM^{P}_{ij} + FAM^{Q}_{ij}}
\end{equation}

\vspace{0.3cm}

\textbf{Step 5:} The encoded fused image $E^F_{ij}$ is acquired by adding the encoded features of the two input images after convolution with the individual weight maps as follows;
\begin{equation}
    E^F_{ij} = W^{P}_{ij} * E^P + W^{Q}_{ij} * E^Q
\end{equation}

\vspace{0.3cm}

\textbf{Step 6:} Finally, generate the fused image $F = F_{i,j}$ by passing the fused encoded image to the decoder ($D$) of the trained CAE model.
\begin{equation}
    F = D(E^F_{ij})
\end{equation}
In the case of the color images (SPECT/PET), the color fused image is generated by performing color space conversion from \textit{YUV} to \textit{RGB} using the fused image $F$ and $U_c$, $V_c$ channels of the original input color images. The color space conversion is performed by using the following equation:
\begin{equation}
\begin{bmatrix} 
R_c \\
G_c \\
B_c \\
\end{bmatrix} =
\begin{bmatrix} 
1 & 0 & 1.14 \\
1 & $- 0.39$ & 0.58 \\
1 & 2.03 & 0 \\
\end{bmatrix}
\begin{bmatrix} 
F \\
U_c \\
V_c \\
\end{bmatrix}   
\end{equation}
\vspace{0.2cm}
%%%%%%%%% SMALLER Figure Fusion Strategy Image %%%%%%% 
% \begin{figure}[hbt!]
% \centerline{\includegraphics[width=.5\textwidth, height=0.23\textwidth]{Image Fusion Framework}}
% \caption{Image fusion framework using the trained model.}
% \label{fig:FusionFramework}
% \end{figure}
% \FloatBarrier 

\section{Experimental Details } \label{section:4}
\subsection{Dataset} 
The proposed CAE model is trained and tested on a  dataset of a total of 2211 images from the Harvard Medical School's whole brain atlas, consisting of CT-MR T2, SPECT-MR T2, and PET-MR T2 pairs, which are considered to generate the training dataset {\cite{summers2003harvard}. The dataset is created by dividing each original image of spatial size \begin{math} 256 \times 256 \end{math} into four image patches of size \begin{math} 64 \times 64 \end{math}. Some patches having no or less information were discarded, and the dataset with 6756 image patches was used to train the model. The test dataset containing 100 image pairs of MR-CT, MR-SPECT, and MR-PET scans of patients with diverse neurological afflictions is used to validate the capability of the proposed method for the fusion of a range of medical images with multiple imaging modalities.
\subsection{Parameter Settings and Implementation Details} 
The proposed model is trained for epochs = 30, batch size = 32, and learning rate \begin{math} = 1 \times  10^{-3} \end{math} using  an  Adam\cite{kingma2014adam} optimizer. All these hyperparameters are set empirically. The loss function used is the mean square error (MSE) between the input image and the reconstructed image. The implementation was performed in the PyTorch framework, using the hardware platform with Intel Core Xeon(R) Silver 4210R CPU, 2.4 GHz, 128 GB RAM, and 24 GB NVIDIA GPU with Ubuntu 22.04 LTS 64-bit operating system. 
 
 A detailed comparative study between the subjective and quantitative results of the proposed and existing state-of-the-art (SOTA) fusion methods is performed to evaluate the effectiveness of the proposed fusion method. The SOTA fusion methods considered are dual-discriminator conditional GANs-based methods by Ma \textit{et al.} (DDcGAN)\cite{ma_ddcgan:_2020}, dual-stream attention mechanism based method by Fu \textit{et al.} (DSAGAN)\cite{fu_dsagan:_2021}, a squeeze-decompose network based method by Zhang \textit{et al.} (SDNet)\cite{zhang_sdnet:_2021}, a dense net based unified fusion framework by Xu \textit{et al.} (U2Fusion)\cite{xu_u2fusion:_2020}. Nine SOTA fusion metrics are used for quantitative evaluation as listed below,
\begin{enumerate}
    \item Entropy ($EN$) \cite{das_nsst_2020}
    \item Standard Deviation ($SD$) \cite{das_nsst_2020}
    \item Spatial Frequency ($SF$) \cite{das_nsst_2020}
    \item Edge Preservation Index ($Q_{AB/F}$) \cite{xydeas_objective_2000}
    \item Mutual Information ($MI$) \cite{qu_information_2002}
    \item Cvejic's metric ($Q_{C}$)\cite{cvejic_similarity_2005}
    \item Yang's metric ($Q_{Y} $)\cite{li_novel_2008}
    \item Sum of the Correlations of Differences ($SCD$) \cite{aslantas_new_2015}
    \item Visual Information Fidelity for Fusion ($VIFF$) \cite{han_new_2013}.
\end{enumerate}

\section{Results and Discussions } \label{section:5}
\subsection{Visual Performance Analysis}
\figurename{~\ref{fig:zoomSOTAResultImage}} provides the visual comparison of the results obtained with SOTA fusion methods and the proposed method. \figurename{~\ref{fig:zoomSOTAResultImage}} (a) and (b) shows the source MR and CT/SPECT/PET images, respectively. Fusion results of the SOTA methods and the proposed method are shown in (c)-(g). From \figurename{~\ref{fig:zoomSOTAResultImage}}(g), it can be seen that the fused images obtained by the proposed fusion method have strong contrast and provide good visualization of information present in both grey and color source images. In \figurename{~\ref{fig:zoomSOTAResultImage}}, zoomed regions are also highlighted by red and green squares for better visualization of the fusion results.

The fused images obtained by the DDcGAN\cite{ma_ddcgan:_2020} method are shown in \figurename{~\ref{fig:zoomSOTAResultImage}}(c). For MR T2-CT pairs 1 and 2, the DDcGAN\cite{ma_ddcgan:_2020}  fails to conserve the structural information in the source images with appropriate contrast. The loss of vital diagnostic edge information is also observed in the fusion results. As compared to the proposed approach, it fails to adequately preserve the soft tissue demarcation present in the MR T2 source images. The fusion results of DSAGAN\cite{fu_dsagan:_2021} are shown in \figurename{~\ref{fig:zoomSOTAResultImage}}(d). For grey image pairs 1 and 2, the soft tissue structure in MR T2 images is sufficiently preserved, but compared to the proposed method, the fusion results cease to preserve the significant demarcation of the skull boundaries present in CT images. For image pairs 3-6, the color information is adequately retained in the fused images, but the contrast of tissue composition in MR T2 images is not retained well. In the case of fusion methods SDNet\cite{zhang_sdnet:_2021} and U2Fusion\cite{xu_u2fusion:_2020}, for MR T2-CT image pairs 1 and 2, the fusion results capture the demarcation of tissue structure in MR T2 images but fail to produce the contrast of the hard tissue composition in CT images. In the case of MR T2-SPECT/PET image pairs 3-6, both SDNet\cite{zhang_sdnet:_2021} and U2Fusion\cite{xu_u2fusion:_2020} methods lack color consistency as compared to the proposed method and the bright pixels look much darker as compared to the original MR T2 images.
 
Compared to the SOTA fusion methods, the visual differences in the results of the proposed fusion method are very prominent, as it proficiently highlights the global as well as local contrast of the six input image pairs. For MR T2-CT pairs 1 and 2, the fused images obtained with the proposed method provide better delineation of both soft and hard tissue structure compared to the SOTA methods. The fused images obtained with the proposed method effectively preserve the complementary information present in the source images and provide rich color consistency for multimodal color image pairs 3-6. In the fused outcomes of the proposed method in \figurename{~\ref{fig:zoomSOTAResultImage}}(g), the edges are also clearly visible along with the structural and textural composition of the source images. The proposed approach provides better visualization of the source images in the fusion results, which may assist medical professionals in making an accurate clinical diagnosis in an efficient manner.  Overall, from the experimental visual results, the proposed method outperforms the SOTA methods in terms of the amount of local and global feature details transferred from the source images to the fused image. 

% %%%%%%%% SOTA Quantitative Analysis Figure %%%%%%%
% \begin{figure*}[!hbtp]
% \centerline{\includegraphics[width=0.8\textwidth, height=0.7\textwidth]{New SOTA Fusion Methods images}}
% \caption{Subjective Comparison of fusion results for State-of-the-art (SOTA) and proposed fusion method. (a) Input Image 1 (MR T2 Image), (b) Input Image 2 (CT/SPECT/PET Image), (c) DDcGAN \cite{ma_ddcgan:_2020}, (d) DSAGAN \cite{fu_dsagan:_2021}, (e) SDNet \cite{zhang_sdnet:_2021}, (f) U2Fusion \cite{xu_u2fusion:_2020}, (g) Proposed Fusion Method}
% \label{fig:SOTAResultImage}
% \end{figure*}
\subsection{Quantitative Performance Analysis}

The quantitative results for image pairs in \figurename{~\ref{fig:zoomSOTAResultImage}} are presented in \tablename{~\ref{table:Qualitative-Individual-SOTA}}.  For every image pair in \tablename{~\ref{table:Qualitative-Individual-SOTA}}, the bold text highlights the best-performing method, while underlined values represent the second-best method. The proposed method shows a significant increment for $EN$ metric for grey-scale and color multimodal image fusion which shows the amount of information present in the fusion results of the proposed approach significantly higher than in the aforementioned SOTA fusion methods. The $SD$ metric, which measures the quality of contrast of fused images compared to the individual source images, is higher for the proposed method. This implies that the fusion outcomes of the proposed method provide stronger contrast compared to other methods. The mutual information content present in fused images is measured by the $MI$ metric. The proposed approach attains the highest values for $MI$ values for most of the image pairs. The proposed method generates fusion results with minimum artifacts and distortion as compared to the other methods as higher values are observed for the visual fidelity $VIFF$ metric for most image pairs. The proposed method performs second best for the $Q_{Y}$ metric, which measures the amount of complementary information preserved. In the case of performance metrics such as $SCD$, $Q_{C}$ and $Q_{AB/F}$, SDNet\cite{zhang_sdnet:_2021} attains higher values for color image pairs 3-6. But the proposed method achieves better results for grey-scale image pairs 1 and 2. Despite the higher objective values gained by SDNet\cite{zhang_sdnet:_2021}, the subjective results obtained by it are secondary to the results of the proposed method. In comparison, the proposed method accurately emulates the contrast of the source images, which is also validated by the higher values of the standard deviation $SD$ metric. Especially for the fusion of MR-CT images, the quantitative and the subjective assessments display significant appeal for the fusion results obtained with the proposed method.

%%%%%%%% Zoomed results SOTA Quantitative Analysis Figure %%%%%%%
\begin{figure*}[!hbtp]
\centering
	\scalebox{0.85}{\includegraphics[width=7.4in, %height=3.4in
		]{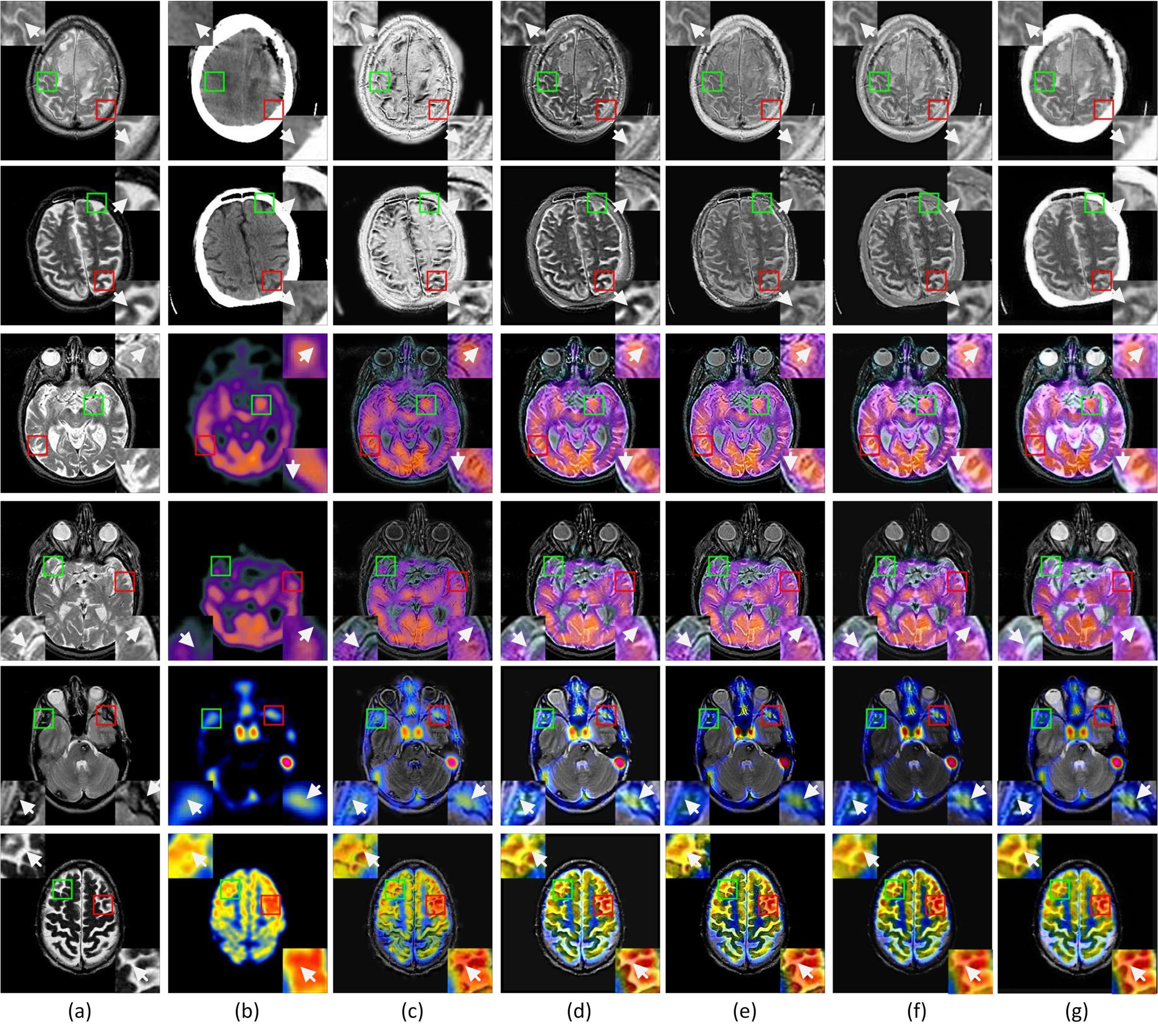}}
\caption{Subjective Comparison of fusion results for State-of-the-art (SOTA) and proposed fusion method. (a) Input Image 1 (MR T2 Image), (b) Input Image 2 (CT/SPECT/PET Image), (c) DDcGAN \cite{ma_ddcgan:_2020}, (d) DSAGAN \cite{fu_dsagan:_2021}, (e) SDNet \cite{zhang_sdnet:_2021}, (f) U2Fusion \cite{xu_u2fusion:_2020}, (g) Proposed fusion method}
\label{fig:zoomSOTAResultImage}
\end{figure*}

%%%%%%%% SOTA Quantitative Analysis Img-Table %%%%%%%
\begin{table*}
\centering
\caption{Quantitative performance of fusion methods for \figurename{~\ref{fig:zoomSOTAResultImage}}}
\label{table:Qualitative-Individual-SOTA}
\begin{tabular}{llccccccccc}
\hline
\multicolumn{1}{l}{Image Pair} & Fusion Methods  & \multicolumn{1}{c}{\textit{EN}} & \multicolumn{1}{c}{\textit{SD}} & \multicolumn{1}{c}{\textit{SF}} & \multicolumn{1}{c}{\textit{\begin{math} Q_{AB/F} \end{math}}} & \multicolumn{1}{c}{\textit{MI}} & \multicolumn{1}{c}{\textit{\begin{math} Q_{C} \end{math}}} & \multicolumn{1}{c}{\textit{\begin{math} Q_{Y} \end{math}}} & \multicolumn{1}{c}{\textit{SCD}} & \multicolumn{1}{c}{\textit{VIFF}} \\ \hline
\multirow{5}{*}{Pair 1} & DDcGAN \cite{ma_ddcgan:_2020} & 5.259 & {\ul 88.883} & \textbf{8.441} & 0.27 & 2.663 & 0.519 & 0.438 & 1.39 & 0.33 \\
& DSAGAN \cite{fu_dsagan:_2021} & {\ul 5.373} & 57.505 & {\ul 7.989} & 0.348 & 2.779 & {\ul 0.606} & 0.587 & 1.082 & 0.25 \\
& SDNet \cite{zhang_sdnet:_2021} & 4.753 & 69.905 & 6.976 & 0.399 & {\ul 3.156} & 0.577 & 0.58 & {\ul 1.391} & 0.376 \\
& U2Fusion \cite{xu_u2fusion:_2020} & 4.623 & 66.876 & 7.156 & \textbf{0.52}  & 2.996 & \textbf{0.625} & \textbf{0.661} & 1.359 & {\ul 0.441} \\
& Proposed Method & \textbf{5.881} & \textbf{90.002} & 6.466 & {\underline{0.405}} & \textbf{3.267} & 0.582 & {\underline{0.609}} & \textbf{1.459} & \textbf{0.56} \\ \hline
\multirow{5}{*}{Pair 2} & DDcGAN \cite{ma_ddcgan:_2020} & 5.21 & \textbf{89.66}  & \textbf{8.048} & 0.255 & 2.756 & 0.491 & 0.365 & {\ul 1.457} & 0.277 \\
& DSAGAN \cite{fu_dsagan:_2021} & {\ul 5.372} & 58.079 & {\ul 7.925} & 0.361 & 2.954 & \textbf{0.63} & {\ul 0.576} & 1.356 & 0.199 \\
& SDNet \cite{zhang_sdnet:_2021} & 4.65 & 53.994 & 7.062 & 0.334 & {\ul 3.077} & {\ul 0.6} & 0.55 & 1.409 & 0.221 \\
& U2Fusion \cite{xu_u2fusion:_2020} & 4.588 & 53.006 & 6.815 & {\ul 0.393} & 3.067 & 0.564 & 0.568 & 1.32 & {\ul 0.302}\\
& Proposed Method & \textbf{5.843} & {\underline{81.92}} & 6.519 & \textbf{0.4} & \textbf{3.367} & 0.551 & \textbf{0.592} & \textbf{1.515} & \textbf{0.435} \\ 
\hline
\multirow{5}{*}{Pair 3} & DDcGAN \cite{ma_ddcgan:_2020} & {\ul 6.254} & 57.921 & 8.278 & 0.418  & 2.821 & 0.607 & 0.536 & 0.788 & 0.212 \\
& DSAGAN \cite{fu_dsagan:_2021} & 5.475 & 71.827 & {\ul 8.842} & 0.548 & 3.105 & {\ul 0.75} & {\ul 0.744} & {\ul 1.271} & 0.421 \\
& SDNet \cite{zhang_sdnet:_2021} & 5.396 & 73.22 & \textbf{9.37} & \textbf{0.64}  & 3.166 & \textbf{0.814} & \textbf{0.832} & \textbf{1.381} & 0.44 \\
& U2Fusion \cite{xu_u2fusion:_2020} & 5.733 & {\ul 73.221} & 8.494 & {\ul 0.609} & {\ul 3.333} & 0.661 & 0.634 & 0.713 & \textbf{0.486} \\
& Proposed Method & \textbf{6.292} & \textbf{83.962} & 8.553 & 0.51 & \textbf{3.471} & 0.693 & 0.644 & 0.919 & \textbf{0.486} \\ 
\hline
\multirow{5}{*}{Pair 4} & DDcGAN \cite{ma_ddcgan:_2020} & {\ul 6.009} & 55.736 & 8.195 & 0.479 & 2.675 & 0.62 & 0.589 & 0.893 & 0.278 \\
& DSAGAN \cite{fu_dsagan:_2021} & 5.444 & {\ul 69.396} & {\ul 8.654} & {\ul 0.586} & 2.903 & {\ul 0.788} & {\ul 0.794} & {\ul 1.573} & 0.527  \\
& SDNet \cite{zhang_sdnet:_2021} & 5.337 & 68.964 & \textbf{9.055} & \textbf{0.651} & 3.014 & \textbf{0.819} & \textbf{0.841} & \textbf{1.595} & \textbf{0.531} \\
& U2Fusion \cite{xu_u2fusion:_2020} & 5.568 & 63.574 & 7.956 & 0.521 & {\ul 3.107} & 0.63 & 0.624 & 0.977 & {\ul 0.528}  \\
& Proposed Method & \textbf{6.157} & \textbf{71.092} & 8.245  & 0.524 & \textbf{3.132} & 0.694 & 0.667 & 0.826 & 0.478 \\ 
\hline
\multirow{5}{*}{Pair 5} & DDcGAN \cite{ma_ddcgan:_2020} & 4.917 & 51.242 & {\ul 7.215} & 0.456 & 2.106 & 0.602 & 0.544 & 1.504 & 0.292 \\
& DSAGAN \cite{fu_dsagan:_2021} & \textbf{5.33} & \textbf{60.022} & \textbf{7.424} & {\ul 0.564} & 2.408 & 0.692 & 0.652 & \textbf{1.82}  & \textbf{0.517} \\
& SDNet \cite{zhang_sdnet:_2021} & 4.02 & 48.495 & 6.688 & \textbf{0.608} & \textbf{2.669} & \textbf{0.777} & \textbf{0.772} & {\ul 1.81} & 0.341  \\
& U2Fusion \cite{xu_u2fusion:_2020} & 4.29 & 41.843 & 5.917 & 0.47 & {\ul 2.512} & 0.618 & 0.548 & 1.693 & 0.343 \\
& Proposed Method & {\underline{ 4.997}} & {\underline{54.125}} & 6.532 & 0.501 & 2.444 & {\underline{0.703}} & {\underline{0.674}} & 1.678 & {\underline{0.393}} \\ 
\hline
\multirow{5}{*}{Pair 6} & DDcGAN \cite{ma_ddcgan:_2020} & 3.85 & 56.12 & 6.775 & 0.339 & 2.57 & 0.581 & 0.548 & 0.766 & 0.247 \\
& DSAGAN \cite{fu_dsagan:_2021} & \textbf{4.346} & {\ul 61.509} & {\ul 7.109} & 0.455 & 2.654 & 0.676 & 0.627 & 1.262 & 0.409 \\
& SDNet \cite{zhang_sdnet:_2021} & 3.222  & 61.262 & \textbf{7.151} & \textbf{0.514} & {\ul 2.784} & \textbf{0.746} & \textbf{0.736} & \textbf{1.568} & \textbf{0.474} \\
& U2Fusion \cite{xu_u2fusion:_2020} & 3.409 & 56.527 & 6.129 & {\ul 0.477} & \textbf{2.801} & 0.647 & 0.607 & {\ul 1.363} & 0.44 \\
& Proposed Method & {\underline{4.185}} & \textbf{61.627} & 6.557 & 0.404 & 2.688 & {\underline{0.679}} & {\underline{0.651}} & 1.358 & {\underline{0.444}} \\ 
\hline
\end{tabular}
\end{table*}
%%%%%%%% SOTA Average Performance TABLE %%%%%%%
\phantom{~}\noindent
\begin{table*}
\centering
\caption{Averaged performance analysis for SOTA fusion methods (\begin{math} \mbox{average} \pm \mbox{standard deviation} \end{math}) }
\label{table:Qualitative-Average-SOTA}
\scalebox{0.85}{%
\begin{tabular}{llllllllll}
\cline{1-10}
Fusion Method & \multicolumn{1}{c}{$EN$} & \multicolumn{1}{c}{$SD$} & \multicolumn{1}{c}{$SF$} & \multicolumn{1}{c}{$Q_{AB/F}$} & \multicolumn{1}{c}{\textit{MI}} & \multicolumn{1}{c}{$Q_{C}$} & \multicolumn{1}{c}{$Q_{Y}$} & \multicolumn{1}{c}{$SCD$} & \multicolumn{1}{c}{$V IF F$}  \\ \cline{1-10}
DDcGAN \cite{ma_ddcgan:_2020} & \multicolumn{1}{c}{5.395 $\pm$ 0.627} & \multicolumn{1}{c}{74.144 $\pm$  3.696} & \multicolumn{1}{c}{7.805 $\pm$  0.506} & \multicolumn{1}{c}{0.372 $\pm$  0.055} & \multicolumn{1}{c}{2.734 $\pm$  0.220} & \multicolumn{1}{c}{0.551 $\pm$  0.035} & \multicolumn{1}{c}{0.494 $\pm$  0.045} & \multicolumn{1}{c}{1.275 $\pm$ 0.219} & \multicolumn{1}{c}{0.266 $\pm$  0.042} \\ 
DSAGAN \cite{fu_dsagan:_2021} & \multicolumn{1}{c}{5.410 $\pm$  0.379} & \multicolumn{1}{c}{60.954 $\pm$  2.898} & \multicolumn{1}{c}{7.900 $\pm$  0.525} & \multicolumn{1}{c}{0.450 $\pm$  0.044} & \multicolumn{1}{c}{2.845 $\pm$  0.256} & \multicolumn{1}{c}{0.658 $\pm$  0.049} & \multicolumn{1}{c}{0.633 $\pm$  0.057} & \multicolumn{1}{c}{1.303 $\pm$  0.188} & \multicolumn{1}{c}{0.333 $\pm$  0.073} \\ 
SDNet \cite{zhang_sdnet:_2021} & \multicolumn{1}{c}{4.682 $\pm$  0.529} & \multicolumn{1}{c}{61.700 $\pm$  6.345} & \multicolumn{1}{c}{7.448 $\pm$  0.680} & \multicolumn{1}{c}{0.496 $\pm$  0.044} & \multicolumn{1}{c}{3.052 $\pm$  0.247} & \multicolumn{1}{c}{0.690 $\pm$  0.037} & \multicolumn{1}{c}{0.693 $\pm$  0.044} & \multicolumn{1}{c}{1.484 $\pm$  0.117} & \multicolumn{1}{c}{0.370 $\pm$  0.079} \\ 
U2Fusion \cite{xu_u2fusion:_2020} & \multicolumn{1}{c}{4.744 $\pm$  0.546} & \multicolumn{1}{c}{56.867 $\pm$  6.788} & \multicolumn{1}{c}{6.97 $\pm$  0.569} & \multicolumn{1}{c}{0.467 $\pm$  0.046} & \multicolumn{1}{c}{2.973 $\pm$  0.258} & \multicolumn{1}{c}{0.586 $\pm$  0.051} & \multicolumn{1}{c}{0.592 $\pm$  0.056} & \multicolumn{1}{c}{1.274 $\pm$ 0.222} & \multicolumn{1}{c}{0.393 $\pm$  0.076} \\ 
Proposed Method & \multicolumn{1}{c}{5.696 $\pm$  0.487} & \multicolumn{1}{c}{71.314 $\pm$  6.513} & \multicolumn{1}{c}{6.786 $\pm$  0.642} & \multicolumn{1}{c}{0.499 $\pm$  0.046} & \multicolumn{1}{c}{3.064 $\pm$  0.303} & \multicolumn{1}{c}{0.596 $\pm$  0.059} & \multicolumn{1}{c}{0.592 $\pm$  0.050} & \multicolumn{1}{c}{1.387 $\pm$  0.228} & \multicolumn{1}{c}{0.409 $\pm$  0.078} \\ \cline{1-10}
\end{tabular}}
\end{table*}

To present a more concise analysis of the aforementioned fusion methods, the averaged quantitative performance is presented in \tablename{~\ref{table:Qualitative-Average-SOTA}} and the following observations are drawn,
\begin{enumerate}
    \item The proposed method achieves 5.58\%, 5.29\%, 21.66\% and 20.07\% greater values of $EN$ than the methods mentioned in \tablename{~\ref{table:Qualitative-Individual-SOTA}} DDcGAN\cite{ma_ddcgan:_2020}, DSAGAN\cite{fu_dsagan:_2021}, SDNet\cite{zhang_sdnet:_2021} and U2Fusion\cite{xu_u2fusion:_2020}, respectively. The greater entropy values show that, in comparison to other approaches, the proposed method's fusion outcomes have rich information content, which validates the visual performance analysis.
    \item In comparison to the methods DDcGAN\cite{ma_ddcgan:_2020}, DSAGAN\cite{fu_dsagan:_2021}, SDNet\cite{zhang_sdnet:_2021} and U2Fusion\cite{xu_u2fusion:_2020}, for $MI$ metric, the proposed method provides 12.07\%, 7.7\%, 0.39\% and 3.06\% higher values, respectively. The fusion outcomes of the proposed approach are improved by the overall higher extent of aggregated information captured from the source images.
    \item The proposed method yields 53.76\%, 22.82\%, 10.54\% and 4.07\% higher values of $VIFF$ than the aforementioned methods DDcGAN\cite{ma_ddcgan:_2020}, DSAGAN\cite{fu_dsagan:_2021}, SDNet\cite{zhang_sdnet:_2021} and U2Fusion\cite{xu_u2fusion:_2020}, respectively. It implies that the proposed approach surpasses other fusion methods for the visual fidelity index and provides fusion results that are visually consistent, without distortion and artifacts.
    \item For $SD$ measure, the proposed method attains 17\%, 15.58\%, and 25.4\% higher values for the methods DSAGAN\cite{fu_dsagan:_2021}, SDNet\cite{zhang_sdnet:_2021} and U2Fusion\cite{xu_u2fusion:_2020}, respectively which denotes that the proposed method's fusion outcomes have superior contrast than the earlier mentioned methods.  
    \item The proposed method gets 2.51\%, 0.31\%, and 2.59\% greater values of $SCD$ than the methods DDcGAN\cite{ma_ddcgan:_2020}, DSAGAN\cite{fu_dsagan:_2021} and U2Fusion\cite{xu_u2fusion:_2020}, respectively. This implies that the proposed method's fusion outcomes have a higher correlation with the reference images and the complementary information is also well-preserved. 
\end{enumerate}

\subsection{Ablation Studies:} 
To explore the effectiveness of the LGCA pooling technique in enhancing the capability of the proposed fusion approach,  an ablation study is conducted with the two most common techniques i.e. average and max pooling. The visual comparison is presented in \figurename{~\ref{fig:zoomAblationResultImages}}. Here, \figurename{~\ref{fig:zoomAblationResultImages}} (a) and (b) display the source images and \figurename{~\ref{fig:zoomAblationResultImages}} (c), (d), and (e) display the fusion results obtained by employing the average, max, and LGCA  pooling techniques, respectively, in the CAE architecture used in the proposed fusion method. The rest of the blocks of the CAE, dataset,  hyperparameters, and implementation platform are kept the same. \figurename{~\ref{fig:zoomAblationResultImages}} shows specific zoomed regions of the fused images for better visualization of the results obtained using each pooling technique. 

 %%%%%%%% Ablation study figure with zoomed regions %%%%%%%
\begin{figure}[!hbt] 
\centering 
\scalebox{0.45}{\includegraphics[width=7.4in, %height=3.4in 
]{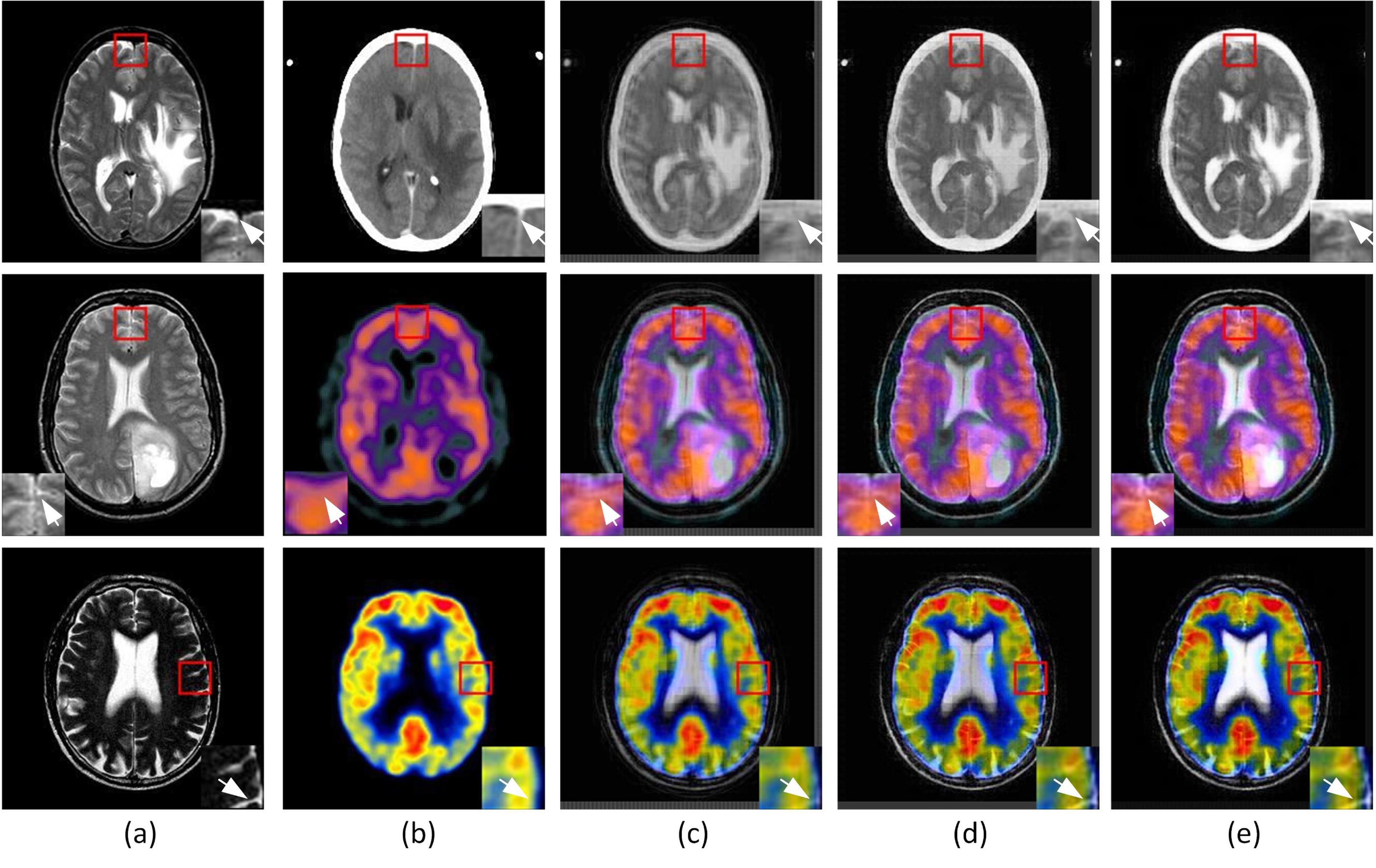}}
\caption{Subjective Comparison of fusion results with three pooling methods (a) Input Image 1 (MR T2 Image), (b) Input Image 2 (CT/SPECT/PET Image), (c) Average pooling, (d) Max pooling, (e) LGCA pooling.}
\label{fig:zoomAblationResultImages}
\end{figure}

It can be observed from  \figurename{~\ref{fig:zoomAblationResultImages}} that the fusion results of each pooling technique capture the structural information accurately but for the average pooling in \figurename{~\ref{fig:zoomAblationResultImages}} (c), the fused images are of very low resolution with muted edges and blurred spatial details of the input images. Efficient preservation of the edge information with vivid contrast is observed in the fusion results of the LGCA pooling. For MR T2-CT fusion, compared to the max pool fused image, both soft and hard tissue demarcations with strong contrast are produced in the fusion result of LGCA pooling. For MR T2-SPECT and MR T2-PET fusion, compared to the max pool, LGCA pooling fusion results are of richer color consistency and finer local as well as global contrast. The soft tissue delineation along the skull's inner boundary in MR images is also more prominent in the resultant fused images with the LGCA pooling method. Compared to the fusion results of the other pooling techniques, the complementary information is preserved to a greater extent by the LGCA pooling.

%%%%%%%% Ablation individual Img-Table %%%%%%%
\begin{table*}
\centering
\caption{Quantitative performance of the pooling methods for \figurename{~\ref{fig:zoomAblationResultImages}}}
\label{table:Qualitative-Individual-Ablation}
\begin{tabular}{llllllllllllll}
\cline{1-11}
Image Pair & Pooling Methods & \multicolumn{1}{c}{\textit{EN}} & \multicolumn{1}{c}{\textit{SD}} & \multicolumn{1}{c}{\textit{SF}} & \multicolumn{1}{c}{\textit{\begin{math} Q_{AB/F} \end{math}}} & \multicolumn{1}{c}{\textit{MI}} & \multicolumn{1}{c}{\textit{\begin{math} Q_{C} \end{math}}} & \multicolumn{1}{c}{\textit{\begin{math} Q_{Y} \end{math}}} & \multicolumn{1}{c}{\textit{SCD}} & \multicolumn{1}{c}{\textit{VIFF}} \\ \cline{1-11}
\multirow{3}{*}{Pair 1} & Average & \multicolumn{1}{c}{5.808} & \multicolumn{1}{c}{65.671} & \multicolumn{1}{c}{{\ul 6.034}} & \multicolumn{1}{c}{0.247} & \multicolumn{1}{c}{3.177} & \multicolumn{1}{c}{0.511} & \multicolumn{1}{c}{0.489} & \multicolumn{1}{c}{1.220} & \multicolumn{1}{c}{0.362} \\
& Max & \multicolumn{1}{c}{\textbf{6.015}} & \multicolumn{1}{c}{{\ul 71.339}} & \multicolumn{1}{c}{5.783} & \multicolumn{1}{c}{{\ul 0.371}}  & \multicolumn{1}{c}{\textbf{3.447}} & \multicolumn{1}{c}{{\ul 0.558}} & \multicolumn{1}{c}{{\ul0.586}} & \multicolumn{1}{c}{{\ul 1.319}} & \multicolumn{1}{c}{{\ul 0.429}}\\
& LGCA & \multicolumn{1}{c}{{\underline{5.882}}} & \multicolumn{1}{c}{\textbf{78.805}} & \multicolumn{1}{c}{\textbf{6.186}} & \multicolumn{1}{c}{\textbf{0.420}}  & \multicolumn{1}{c}{{\underline{3.436}}} & \multicolumn{1}{c}{\textbf{0.589}} & \multicolumn{1}{c}{\textbf{0.630}} & \multicolumn{1}{c}{\textbf{1.464}} & \multicolumn{1}{c}{\textbf{0.448}} \\ \cline{1-11}
\multirow{3}{*}{Pair 2} & Average & \multicolumn{1}{c}{\textbf{6.163}} & \multicolumn{1}{c}{66.467} & \multicolumn{1}{c}{6.670}  & \multicolumn{1}{c}{0.222} & \multicolumn{1}{c}{3.034} & \multicolumn{1}{c}{0.530} & \multicolumn{1}{c}{0.488} & \multicolumn{1}{c}{0.637} & \multicolumn{1}{c}{0.443} \\
& Max & \multicolumn{1}{c}{{\ul 6.130}} & \multicolumn{1}{c}{{\ul 66.789}} & \multicolumn{1}{c}{{\ul 6.795}} & \multicolumn{1}{c}{{\ul 0.407}} & \multicolumn{1}{c}{{\ul 3.267}} & \multicolumn{1}{c}{{\ul 0.617}} & \multicolumn{1}{c}{{\ul 0.600}}  & \multicolumn{1}{c}{{\ul 0.650}} & \multicolumn{1}{c}{\textbf{0.503}}  \\
& LGCA & \multicolumn{1}{c}{5.997} & \multicolumn{1}{c}{\textbf{70.970}}  & \multicolumn{1}{c}{\textbf{6.883}} & \multicolumn{1}{c}{\textbf{0.468}} & \multicolumn{1}{c}{\textbf{3.373}} & \multicolumn{1}{c}{\textbf{0.681}} & \multicolumn{1}{c}{\textbf{0.669}} & \multicolumn{1}{c}{\textbf{0.871}} & \multicolumn{1}{c}{{\underline{0.501}}} \\ \cline{1-11}
\multirow{3}{*}{Pair 3} & Average & \multicolumn{1}{c}{{\ul 5.434}} & \multicolumn{1}{c}{{\ul 65.347}} & \multicolumn{1}{c}{7.079} & \multicolumn{1}{c}{0.220} & \multicolumn{1}{c}{2.653} & \multicolumn{1}{c}{0.457} & \multicolumn{1}{c}{0.526} & \multicolumn{1}{c}{1.511} & \multicolumn{1}{c}{0.308} \\
& Max & \multicolumn{1}{c}{\textbf{5.436}} & \multicolumn{1}{c}{64.780} & \multicolumn{1}{c}{\textbf{7.183}} & \multicolumn{1}{c}{{\ul 0.329}} & \multicolumn{1}{c}{{\ul 2.837}} & \multicolumn{1}{c}{{\ul 0.526}} & \multicolumn{1}{c}{{\ul 0.587}} & \multicolumn{1}{c}{{\ul 1.580}} & \multicolumn{1}{c}{\textbf{0.346}} \\
& LGCA & \multicolumn{1}{c}{5.339} & \multicolumn{1}{c}{\textbf{68.585}} & \multicolumn{1}{c}{{\underline{7.154}}} & \multicolumn{1}{c}{\textbf{0.350}} & \multicolumn{1}{c}{\textbf{2.927}} & \multicolumn{1}{c}{\textbf{0.554}} & \multicolumn{1}{c}{\textbf{0.598}} & \multicolumn{1}{c}{\textbf{1.681}} & \multicolumn{1}{c}{{\underline{0.338}}} \\ \cline{1-11} \end{tabular}
\end{table*}

%%%%%%%% Ablation studies Average Performance TABLE %%%%%%%
\begin{table*}
\centering
\caption{Averaged subjective performance analysis for pooling methods (\begin{math} \mbox{average} \pm \mbox{standard deviation} \end{math}) }
\label{table:Qualitative-Average-Ablation}
\scalebox{0.875}{%
\begin{tabular}{llllllllll}
\hline
Pooling Method & \multicolumn{1}{c}{\textit{EN}} & \multicolumn{1}{c}{\textit{SD}} & \multicolumn{1}{c}{\textit{SF}} & \multicolumn{1}{c}{\textit{\begin{math} Q_{AB/F} \end{math}}} & \multicolumn{1}{c}{\textit{MI}} & \multicolumn{1}{c}{\textit{\begin{math} Q_{C} \end{math}}} & \multicolumn{1}{c}{\textit{\begin{math} Q_{Y} \end{math}}} & \multicolumn{1}{c}{\textit{SCD}} & \multicolumn{1}{c}{\textit{VIFF}} \\ \hline
Average & 5.780 $\pm$  0.460 & 62.709 $\pm$  5.557 & 6.482 $\pm$  0.600 & 0.191 $\pm$  0.022 & 2.763 $\pm$  0.250 & 0.438 $\pm$  0.061 & 0.422 $\pm$  0.050 & 1.025 $\pm$  0.251 & 0.344 $\pm$  0.06 \\ 
Max & 5.893 $\pm$  0.445 & 66.449 $\pm$  4.935 & 6.558 $\pm$  0.641 & 0.338 $\pm$  0.038 & 3.014 $\pm$  0.283 & 0.525 $\pm$  0.047 & 0.535 $\pm$  0.039 & 1.104 $\pm$  0.269 & 0.400 $\pm$  0.067 \\ 
LGCA & 5.696 $\pm$  0.487 & 71.314 $\pm$  6.513 & 6.786 $\pm$  0.642 & 0.399 $\pm$  0.046 & 3.064 $\pm$  0.303 & 0.566 $\pm$  0.059 & 0.592 $\pm$  0.050 & 1.307 $\pm$  0.228 & 0.409 $\pm$  0.078 \\ \hline
\end{tabular}}
\end{table*}

The quantitative performance analysis for \figurename{~\ref{fig:zoomAblationResultImages}} is presented in \tablename{~\ref{table:Qualitative-Individual-Ablation}}. In \tablename{~\ref{table:Qualitative-Individual-Ablation}}, for each image pair, the LGCA pooling achieves the highest values for performance measures \textit{SD}, \textit{\begin{math} Q_{AB/F} \end{math}}, \textit{\begin{math} Q_{C} \end{math}}, \textit{\begin{math} Q_{Y} \end{math}} and \textit{SCD} which further supports the visual evaluation of the fusion results. Overall for \textit{EN} metric, Max-pool shows significant performance but LGCA-pooling gains notable results for \textit{MI} values. Discussing the visual clarity in fusion outcomes, LGCA-pool has an overall marked increment in \textit{SF} measure. The proposed fusion method achieves top performance for visual information metric \textit{VIFF}, in case of CT-MR fusion, and comes in second to max-pool for the fusion of anatomical-functional pairs 2 and 3 in \tablename{~\ref{table:Qualitative-Individual-Ablation}}. 

The averaged quantitative analysis for the three pooling techniques is presented in \tablename{~\ref{table:Qualitative-Average-Ablation}}. It is observed that the proposed method has significantly better results for metrics \textit{SD}, \textit{SF}, \textit{\begin{math} Q_{AB/F} \end{math}}, \textit{MI}, \textit{\begin{math} Q_{C} \end{math}}, \textit{\begin{math} Q_{Y} \end{math}} and \textit{SCD}. The following observations are drawn,
\begin{enumerate}
    \item The proposed method with LGCA pooling achieves 13.72\% and 7.32\% higher values of $SD$ than the average and max pooling methods, respectively which imply superior contrast as compared to the conventional pooling methods.
    \item The $SF$ performance parameter measures the clarity or sharpness of the fused images. For this metric, the LGCA pooling achieves 4.69\% and 3.48\% greater values compared to the average and max pooling, respectively.
    \item The proposed method with LGCA pooling achieves 10.89\% and 1.66\% higher values of \textit{MI} than the average and max pooling methods, respectively which signify higher visual information.
    \item The $Q_{C}$ metric measures the amount of local similarities preserved with minimum deformations between the source and fused images which exist in the same spatial position. The proposed method with LGCA pooling has 29.22\% and 7.81\% greater values than the average and max pool techniques, respectively.
    \item For $Q_{Y}$ measure, the LGCA pooling achieves 40.28\% and 10.65\% higher values than the average and max pooling, respectively. This shows that, compared to other approaches, the fusion results achieved using LGCA pooling preserve the complementary information of the reference images far better.
    \item In comparison to the average and max pooling, the LGCA pooling method produces $SCD$ values that are 27.51\% and 18.39\% greater, respectively, indicating that the source and fused images achieve higher correlation.
    \item For the edge preservation index $Q_{AB/F}$, the proposed method with LGCA pooling yields 108.9\% and 18.05\% higher values than the average and max pooling, respectively which refers to the amount of edge information.
    \item The LGCA pooling-based proposed model exceeds the results of average and max pooling, respectively, by 18.9\% and 2.25\% for the $VIFF$ metric, indicating the presence of more prominent visual information with minimum distortion and artifacts.
\end{enumerate}

Based on the above observations, it can be observed that the proposed method with the LGCA pooling layer achieves notable improvement in visual and quantitative fusion results as compared to the conventional pooling layers.

\section{Conclusion} \label{section:6}

%Medical image fusion aids in the accurate  diagnosis of diverse clinical afflictions. 
This paper presents a convolutional autoencoder (CAE)-based multimodal medical image fusion method that integrates a Laplacian-Gaussian concatenation with an attention pooling technique as a viable substitute for the conventional pooling methods. The LGCA pooling layers used in the encoder block help in retaining both the high and low-frequency features of the source images leading to the effective preservation of tissue contrast and edges. The weighted average fusion strategy combines the encoded feature maps of the two inputs and helps in the effective reconstruction of the resultant fused image using the trained decoder. Furthermore, training the fusion model on neurological images helps to capture the inherent characteristics of a variety of brain tissues and improves the visualization of the fused images. Detailed performance analysis demonstrates that the proposed fusion method provides fused images with improved visual quality and also outperforms the existing fusion methods in quantitative assessment.

% \newpage
\bibliographystyle{ieeetr}
\bibliography{Main}{} 

\end{document}